\documentclass[pra,twocolumn,superscriptaddress]{revtex4-2}

\usepackage[utf8]{inputenc}
\usepackage[T1]{fontenc}
\usepackage{graphicx}
\usepackage{dcolumn}
\usepackage{bm}
\usepackage{amsmath,amsfonts}
\usepackage{mathtools}
\usepackage{qcircuit}
\usepackage{braket}
\usepackage{color}
\usepackage{times}
\usepackage[normalem]{ulem}
\usepackage{comment}

\usepackage[colorlinks=true,linkcolor=blue,citecolor=blue,breaklinks=true]{hyperref}

\usepackage[capitalize]{cleveref}

\DeclareMathOperator{\Tr}{\mathrm{Tr}}

\begin{document}

\title{Impact of Measurement Noise on Escaping Saddles in Variational Quantum Algorithms}
\author{Eriko Kaminishi}
\affiliation{Quantum Computing Center, Keio University, 3-14-1 Hiyoshi, Kohoku-ku, Yokohama, Kanagawa, 223-8522, Japan}
\author{Takashi Mori}
\affiliation{Department of Physics, Keio University, 3-14-1 Hiyoshi, Kohoku-ku, Yokohama 223-8522, Japan}
\author{Michihiko Sugawara}
\affiliation{Quantum Computing Center, Keio University, 3-14-1 Hiyoshi, Kohoku-ku, Yokohama, Kanagawa, 223-8522, Japan}
\author{Naoki Yamamoto}
\affiliation{Quantum Computing Center, Keio University, 3-14-1 Hiyoshi, Kohoku-ku, Yokohama, Kanagawa, 223-8522, Japan}
\affiliation{Department of Applied Physics and Physico-Informatics, Keio University, 3-14-1 Hiyoshi, Kohoku-ku, Yokohama, Kanagawa, 223-8522, Japan}
\date{\today}
\begin{abstract}
Stochastic gradient descent (SGD) is a frequently used optimization technique in classical machine learning and Variational Quantum Eigensolver (VQE).
For the implementation of VQE on quantum hardware, the results are always affected by measurement shot noise.
However, there are many unknowns about the structure and properties of the measurement noise in VQE and how it contributes to the optimization.
In this work, we analyze the effect of measurement noise to the optimization dynamics.
Especially, we focus on escaping from saddle points in the loss landscape, which is crucial in the minimization of the non-convex loss function. 
We find that the escape time (1) decreases as the measurement noise increases in a power-law fashion and (2) is expressed as a function of $\eta/N_s$ where $\eta$ is the learning rate and $N_s$ is the number of measurements.
The latter means that the escape time is approximately constant when we vary $\eta$ and $N_s$ with the ratio $\eta/N_s$ held fixed.
This scaling behavior is well explained by the stochastic differential equation (SDE) that is obtained by the continuous-time approximation of the discrete-time SGD.
According to the SDE, $\eta/N_s$ is interpreted as the variance of measurement shot noise.
This result tells us that we can learn about the optimization dynamics in VQE from the analysis based on the continuous-time SDE, which is theoretically simpler than the original discrete-time SGD.

\end{abstract}

\maketitle

\section{Introduction}
Quantum computers are a new type of computer capable of very fast computations and play a very important role in modern science and technology. 
However, current Noisy Intermidiate-Scale Quantum (NISQ)~\cite{preskill2018quantum} computers are affected by noise, which requires the development of algorithms for NISQ.
Variational Quantum Algorithms (VQAs) have been developed to address this problem~\cite{Cerezo_2021}.
VQAs are a set of methods for solving optimization problems using quantum computers, which are of particular interest in areas such as quantum chemistry and machine learning VQAs search for optimal solutions to a problem by optimizing the parameters of a quantum circuit. These algorithms employ a hybrid approach of classical and quantum computing and are expected to provide useful applications for the NISQ computer, especially in the fields of chemistry and materials science.

One of the most important examples is VQE (Variational Quantum Eigensolver) \cite{peruzzo2014variational,kandala2017hardware}, an algorithm that uses a quantum computer to obtain the ground state and its energy of a given Hamiltonian. This task is relevant to many quantum chemical problems, such as chemical reactions and the electronic structure of materials.

In VQE, the quantum state $\ket{\psi(\bm{\theta})}$ is parameterized by a set of parameters denoted by $\bm{\theta}$, and $\bm{\theta}$ is optimized so that the loss function $L(\bm{\theta})=\braket{\psi(\bm{\theta})|H|\psi(\bm{\theta})}$, which is nothing but the expectation value of the Hamiltonian $H$, is minimized.
The gradient descent (GD) is one of the simplest algorithm for the optimization.
We make a remark that gradient-free optimizers are also proposed~\cite{nakanishi2020sequential, ostaszewski2021structure, parrish2019jacobi,watanabe2021optimizing}, but this work focuses on the gradient-based methods.
The parameter $\bm{\theta}^{(k)}$ at $k$th step is updated as $\bm{\theta}^{(k+1)}=\bm{\theta}^{(k)}-\eta\nabla L(\bm{\theta}^{(k)})$, where $\eta>0$ is the learning rate.
The gradient information used for parameter updates can be obtained from an infinite number of measurement operations (shots) through the parameter shift rule (see in \cref{noisesgd}). 
However, in practice, the mean value evaluated with finite shots is used, which gives us a stochastic gradient estimator $\hat{g}(\bm{\theta})$.
In the stochastic gradient descent (SGD)~\cite{harrow2021low, tamiya2022stochastic, liu2022noise, Sweke2020stochasticgradient}, the parameter is updated as $\bm{\theta}^{(k+1)}=\bm{\theta}^{(k)}-\eta\hat{g}(\bm{\theta}^{(k)})$~\cite{Sweke2020stochasticgradient}.

The SGD is also frequently used as a powerful technique in classical machine learning, and it has been argued that SGD noise plays a positive role for escaping from local minima and saddle points~\cite{kleinberg2018alternative}, reducing computational cost and fast convergence, as well as for generalization performance.
Those results in machine learning tell us that noisy update of the parameters is not necessarily harmful but rather beneficial for optimization.


In the context of VQA, the finite measurement approach, which corresponds to implementing SGD, offers positive benefits as well. One such benefit is its effectiveness in escaping saddle points and local minima, which commonly appear during the optimization process in VQA.
Those saddle points and local minima make the convergence very slow, or even lead to a wrong solution that differs from a global minimum of the loss function~\cite{Bittel_2021}.
It should be noted that \citet{harrow2021low} showed that SGD can converge faster than other zeroth-order methods at a smaller computational cost when we apply SGD to classical-quantum hybrid algorithms, including VQA.


On the other hand, implementing VQA faces a significant challenge known as the barren plateau problem, making computation on large-scale quantum bits difficult \cite{mcclean2018barren, PRXQuantum.2.040316, CerveroMartin2023barrenplateausin, larocca2024review}. Attempts to circumvent the barren plateau by shallow circuit designs have been observed to make the problem classically simulable \cite{cerezo2024does}. 
Therefore, while substantial research is still required to fundamentally address the barren plateau issue, various strategies such as careful initialization, local observable usage, and shallow circuitry have been proposed as solutions\cite{Grant2019initialization, PhysRevA.106.L060401, cerezo2021cost, PhysRevLett.132.150603, cao2024exploiting, Park2024hamiltonian}. Our study discusses the inevitable effects of measurement shot noise for optimization process when implementing gradient descent on quantum devices. In the context of advancing gradient optimization by avoiding barren plateaus, research on SGD remains crucial, particularly gives insights into gradient-based methods considering the impact of finite measurements noise.

The effect of stochastic noise to the optimization dynamics has been actively studied on recent studies on the classical machine learning.
However, it should be emphasized that the source of noise here differs from that in classical machine learning.
In machine learning, stochastic noise in SGD is introduced by a finite sampling from the training data.
In contrast, we consider noise due to a finite number of measurements of the gradient.
This difference of the source of noise motivates us to understand how stochastic noise in SGD for VQE affects the optimization dynamics.
It is therefore desired to understand roles of stochastic noise in SGD for VQE.

Regarding this problem, we mention that \citet{liu2022noise} proved a relevant result: stochastic noise added to the GD avoids strict saddle points (i.e., saddle points with at least one negative Hessian eigenvalue).
However, in \citet{liu2022noise}, isotropic noise is artificially added to the GD, and hence the actual structure of measurement noise is not taken into account.
Additionally, a quantitative study of the escape time from saddle points, considering measurement noise, has not been conducted.
Our work aims to give a quantitative study on the escape from saddles via measurement noise in VQE.

In this paper, we have obtained the following results in pursuit of the aforementioned objectives.  We find that the escape time is expressed as a function of $\eta/N_s$, and the escape time decreases as $\eta/N_s$ increases in a power-law fashion. Moreover, we show that the behavior of finite-shot SGD can be interpreted using SDE analysis.

The rest of the paper is  organized as follows.
In \cref{sec:noise}, we discuss properties of measurement noise. 
In \cref{sec:escape}, we numerically examine the escape time from a saddle, and investigate how the escape time depends on $N_s$, which denotes the number of quantum measurements in evaluating the loss gradient, and $\eta$, which denotes the learning rate.
In \cref{sec:SDE}, we derive a stochastic differential equation (SDE) as a continuous-time approximation of the SGD.
The SDE is obtained by considering the limit of small learning rate, and some subtlety regarding this limit is also discussed.
The derived SDE correctly reproduces the escape time in \cref{sec:escape}. Remarkably, it well explains the remarkable behavior observed in \cref{sec:escape}: the escape time is expressed as a function of $v^2=\eta/N_s$, where $v$ is interpreted as the noise strength.
In \cref{sec:conclusion}, we conclude our work with some future outlooks.

\section{Measurement noise in variational quantum algorithms}
\label{sec:noise}

\subsection{Setup of gradient-based methods in VQE}
Suppose that we want to obtain the ground state and its energy of a given Hamiltonian $H$ of $n$ qubits.
In VQE, we construct an ansatz state $\ket{\psi(\bm{\theta})}=U(\bm{\theta})\ket{\psi_0}$, where $U(\bm{\theta})$ is an unitary operator corresponding to an entire quantum circuit parameterized by $\bm{\theta}$.
We then try to obtain the ground state of $H$ by minimizing the loss function $L(\bm{\theta})=\braket{\psi(\bm{\theta})|H|\psi(\bm{\theta})}$.
As is briefly mentioned in Introduction, in the GD, the parameters are optimized by repeating the following update:
\begin{align}
    \bm{\theta}^{(k+1)}=\bm{\theta}^{(k)}-\eta\nabla L(\bm{\theta}^{(k)}),
\end{align}
where $\eta>0$ is the learning rate.
However, in principle, infinitely many quantum measurements are necessary to exactly obtain $\nabla L(\bm{\theta})$.
In practice, it is impossible to perform quantum measurement infinitely many times: what we can do is approximately estimating $\nabla L(\bm{\theta})$ by performing a finite number $N_s$ of quantum measurements.
In essence, the gradient calculation is performed by using shot averages.
Because of the stochastic nature of the measurement outcome in quantum theory, the estimator $\hat{g}(\bm{\theta})$ of $\nabla L(\bm{\theta})$ is a stochastic variable.
In the SGD, we optimize $\bm{\theta}$ by using this stochastic estimator:
\begin{align}
    \bm{\theta}^{(k+1)}=\bm{\theta}^{(k)}-\eta\hat{g}(\bm{\theta}^{(k)}).
    \label{eq:SGD}
\end{align}

\subsection{Noise structure of SGD for VQE}\label{noisesgd}
There are several ways to evaluate gradients on a quantum circuit.
Among them, we focus on the parameter-shift rule\cite{mitarai2018quantum, schuld2019evaluating, li2017hybrid}.

Before introducing the parameter-shift rule, we first discuss a naive method called the finite difference method.
The finite difference method gives an approximation of the derivative, whose quality depends on $\epsilon$.
It approximates the gradient as
\begin{equation}\label{eq:FD}
\frac{\partial L(\bm{\theta})}{\partial \bm{\theta}_i}\approx\frac{L(\bm{\theta} + \epsilon\bm{e}_i) - L(\bm{\theta} - \epsilon\bm{e}_i)}{2\epsilon},
\end{equation}
where $\bm{e}_i$ denotes the unit vector along $i$th direction.
In this method, even if we exactly evaluate the right-hand side by performing infinitely many quantum measurements ($N_s\to\infty$), we still have an error due to a finite value of $\epsilon$.
If we choose a too small value of $\epsilon$, the error due to a finite $N_s$ is greatly enhanced.
Thus, there is a trade-off between the error due to a finite $\epsilon$ and that due to a finite $N_s$.

The parameter-shift rule is an unbiased estimator of the gradient, meaning that it gives the exact gradient if we could perform infinitely many quantum measurements.
A crucial assumption is that the unitary gate $U(\theta)$ is generated by a Hermitian operator $G$ that has only two eigenvalues $g_0$ and $g_1$: $U(\theta)=e^{i\theta G}$.
Then, the gradient can be computed by using finite shifts of parameters.
The derivative of the function $f(\theta)=\bra{\psi}U(\theta)OU^\dagger(\theta)\ket{\psi}$ is then given as follows:
\begin{equation}
    \frac{df(\theta)}{d\theta}=r\left[f(\theta+\pi/4r)-f(\theta-\pi/4r)\right],
\end{equation}
where $r=(g_1-g_0)/2$.
All standard parameterized gates that are used in this work have $r=1/2$.
We therefore have
\begin{equation}\label{eq:PS}
\frac{\partial L(\bm{\theta})}{\partial \bm{\theta}_i}= \frac{L(\bm{\theta}+\frac{\pi}{2} \bm{e}_i)-L(\bm{\theta}-\frac{\pi}{2} \bm{e}_i)}{2}.
\end{equation}

\citet{mari2021estimating} compared finite-difference gradient estimator and the parameter-shift gradient estimator: they tested their predictions by numerical simulations and real quantum experiments.
They showed that the error of $\partial L/\partial \theta_i$ in the finite-difference method optimally scales as $N_s^{-1/3}$ by appropriately choosing $\epsilon$ for a given $N_s$, while that of the parameter-shift rule scales as $N_s^{-1/2}$.
The parameter-shift rule thus achieves a smaller error with a fixed number of measurements, so that we utilize parameter-shift rule to implement gradient method in quantum algorithms.

When we use the parameter shift rule as a gradient estimator $\hat{g}(\bm{\theta})$, the SGD dynamics becomes stochastic.
\Cref{eq:SGD} is rewritten as
\begin{equation}
    \bm{\theta}^{(k+1)}=\bm{\theta}^{(k)}-\eta\nabla L(\bm{\theta}^{(k)})+\bm{\xi}(\bm{\theta}^{(k)}),
\end{equation}
where $\bm{\xi}(\bm{\theta})\coloneqq-\eta[\hat{g}(\bm{\theta})-\nabla L(\bm{\theta})]$ is the noise vector.
We denote by $\mathbb{E}[\cdot]$ the average over the infinitely many repetitions of quantum measurements.
We then have $\mathbb{E}[\bm{\xi}]=0$.

In \cref{eq:SGD}, time steps needed for $\bm{\theta}$ to change is proportional to $1/\eta$, during which the total number of measurements is proportional to $N_s/\eta$.
When it is large enough, noise can be regarded as Gaussian, owing to the central limit theorem.
Thus, for small $\eta/N_s$, noise is considered to be Gaussian, which is characterized by the covariance matrix
\begin{equation}
    C(\bm{\theta})\coloneqq \mathbb{E}[\bm{\xi}(\bm{\theta})\bm{\xi}^\top(\bm{\theta})].
\end{equation}
Its matrix element is given by $C_{ij}(\bm{\theta})=\mathbb{E}[\xi_i(\bm{\theta})\xi_j(\bm{\theta})]$.

Now let us calculate the covariance matrix of noise for the parameter-shift rule.
Suppose that the Hamiltonian is decomposed as 
\begin{equation}
    H=\sum_lh_l,
\end{equation} 
and quantum measurements are performed $N_s$ times for each $h_k$ separately.
Under this setting, by using \cref{eq:PS}, $C(\bm{\theta})$ is calculated as
\begin{align}
    C_{ij}(\bm{\theta})&=\frac{\delta_{ij}}{4N_s}\sum_k\left(\bra{ \bm{\theta}^{(i+)}}h_k^2\ket{\bm{\theta}^{(i+)}}-\bra{\bm{\theta}^{(i+)}}h_k\ket{\bm{\theta}^{(i+)}}^2\right. \nonumber\\
    &\left.+\bra{\bm{\theta}^{(i-)}}h_k^2\ket{\bm{\theta}^{(i-)}}
    -\bra{\bm{\theta}^{(i-)}}h_k\ket{\bm{\theta}^{(i-)}}^2\right),
\end{align}
where we used simplified notations $\bm{\theta}^{(i\pm)}\coloneqq\bm{\theta}\pm(\pi/2)\bm{e}_i$.

At this point, we note that when measuring gradients with VQE, each gradient component is evaluated independently. 
The exact gradient $\nabla L(\bm{\theta})$ can be obtained through infinitely many quantum measurements. 
However, in practice, we can only obtain the stochastic gradient estimator given as an average value over a finite number of quantum measurements.
Therefore, the covariance matrix of SGD in VQE has no off-diagonal matrix element: $C_{ij}(\bm{\theta})=0$ for any $i\neq j$.

The structure of the covariance matrix of SGD in VQA differs from that in machine learning, where noise stems from the mini-batch sampling of training data.
In classical machine learning, it was argued that strong anisotropic nature of the noise covariance matrix, which is associated with nontrivial structure of off-diagonal matrix elements, is beneficial for generalization~\cite{wu2018sgd,keskar2016large,hoffer2017train} and for optimization~\cite{jastrzkebski2017three,wu2018sgd,wu2020noisy,zhu2018anisotropic,xie2020diffusion,liu2021noise,mori2022power}.
In contrast, the covariance matrix of SGD for VQE has no off-diagonal elements.

The mini-batch sampling noise in machine learning also has another interesting property: noise strength is correlated with the value of the loss function~\citep{mori2022power}, and noise can vanish at a global minimum.
By contrast, the measurement shot noise does not vanish even if our quantum state $\ket{\psi(\bm{\theta})}$ realizes the exact ground state of $\hat{H}$, unless $\hat{H}$ is a frustration-free Hamiltonian, where all the terms $\{h_k\}$ can be simultaneously minimized in the ground state.

\section{Effect of measurement noise for the optimization dynamics}
\label{sec:escape}

It has been reported that implementing SGD instead of GD makes the convergence faster~\cite{liu2022noise,Sweke2020stochasticgradient}.
Some theorems have shown that noisy update of the parameters can avoid to get stuck at saddle points or local minima and therefore is beneficial for GD optimization in the classical machine learning~\cite{jin2019nonconvex,Jain_2017} and in VQAs\cite{liu2022noise}.
However, theoretical bounds on the escape time are not tight at all, and hence quantitative studies on the escape time from saddles in VQEs are still lacking.
In this section, we consider SGD implemented in VQEs, and numerically evaluate the escape time from saddle points in \cref{sec:saddle} and from an excited state, which is a marginally stable point of the loss function (i.e. the minimum eigenvalue of the Hessian of the loss function is almost zero), in \cref{sec:excited}.

In the following, we consider the problem of finding the ground state of the one-dimensional (1D) Heisenberg model as a prototypical example of VQEs.
The Hamiltonian is given by
\begin{equation}\label{eq:hamiltonian}
H=\sum_{i=1}^4\left( X_iX_{i+1}+Y_iY_{i+1}+Z_iZ_{i+1}\right),
\end{equation}
where $X_i$, $Y_i$, and $Z_i$ are Pauli operators at site $i$.
In performing quantum measurements, we decompose the Hamiltonian as $H=\sum_{l=1}^3h_l$, where
\begin{equation}
    h_1=\sum_{i=1}^4X_iX_{i+1}, \quad h_2=\sum_{i=1}^4 Y_iY_{i+1}, \quad h_3=\sum_{i=1}^4Z_iZ_{i+1}.
\end{equation}
That is, in evaluating $L(\bm{\theta}\pm\pi/2\bm{e}_i)$ in the parameter-shift rule (see \cref{eq:PS}, we perform $N_s$ quantum measurements for each $h_k$ separately.

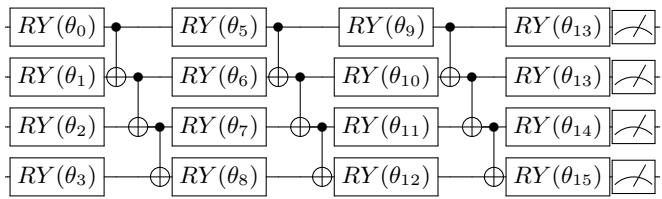
\begin{figure}[tb]
\Qcircuit @C=0.1em @R=0.5em 
{
  & \qw & \gate{RY(\theta_0)} & \ctrl{1}  & \qw  & \qw & \gate{RY(\theta_5)}  & \ctrl{1}  & \qw  & \qw & \gate{RY(\theta_9)} &  \ctrl{1}  & \qw  & \qw & \gate{RY(\theta_{13})} & \meter   & \qw & \qw \\
  & \qw & \gate{RY(\theta_1)} & \targ{} & \ctrl{1} & \qw & \gate{RY(\theta_6)} & \targ{} & \ctrl{1} & \qw & \gate{RY(\theta_{10})} &\targ{} & \ctrl{1} & \qw & \gate{RY(\theta_{13})} &\meter & \qw & \qw \\
  & \qw & \gate{RY(\theta_2)} & \qw & \targ{} & \ctrl{1} \qw & \gate{RY(\theta_7)} & \qw & \targ{} & \ctrl{1} \qw & \gate{RY(\theta_{11})} & \qw & \targ{} & \ctrl{1} \qw & \gate{RY(\theta_{14})} & \meter & \qw & \qw \\
  & \qw & \gate{RY(\theta_3)} & \qw & \qw & \targ{} & \gate{RY(\theta_8)} & \qw & \qw & \targ{} & \gate{RY(\theta_{12})} & \qw & \qw & \targ{} & \gate{RY(\theta_{15})} & \meter & \qw  & \qw \\
}
\caption{4-qubit RY ansatz for 1D Heisenberg model}\label{fgr:ansatz}
\end{figure}

The ansatz is taken as a typical hardware-efficient ansatz that consists of alternating layers of $R_y$ rotations and CNOT entanglements illustrated in Fig. \ref{fgr:ansatz}.
$R_y(\theta)$ denotes the single-qubit rotation operator defined by
\begin{equation}
R_y(\theta)=e^{-i\theta Y/2}=\begin{bmatrix*}
   \mathrm{cos}(\theta/2) & \mathrm{-sin}(\theta/2) \\
   \mathrm{sin}(\theta/2) & \mathrm{cos}(\theta/2)
\end{bmatrix*},
\end{equation}
and CNOT gates between neighboring qubits defined by $|0\rangle\langle0|\otimes I + |1\rangle\langle 1|\otimes X$.

\subsection{Escaping from saddle points}\label{sec:saddle}

In the high-dimensional loss landscape, it is known that most critical points are saddles~\citep{Bray2007,Dauphin2014}, and hence it is significant to efficiently escape from saddles in the gradient-based optimization algorithms.
  
\begin{figure}[tb]
  \centering
  \includegraphics[scale=0.55]{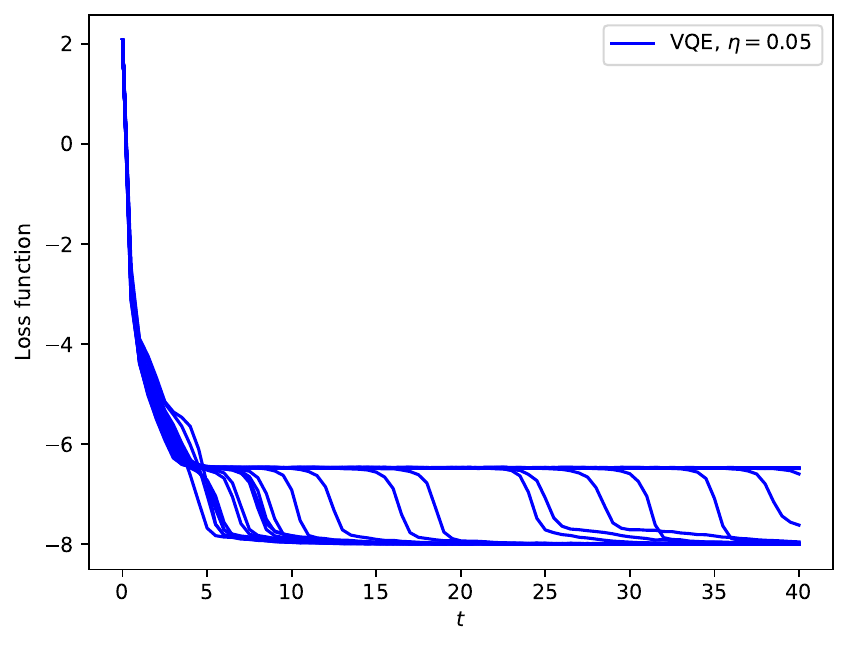}\\
  \caption{Occurrences of saddle points in the SGD convergence process.  We
prepare 30 instances starting from same initial condition.
Learning rate $\eta$ is 0.05.
The number of measurements at each step is 100.
In many cases the parameters $\bm{\theta}$ are trapped at a saddle point before converging to the ground state with the energy $-8$.}\label{fgr:SGDsaddle}
\end{figure}

\Cref{fgr:SGDsaddle} shows the energy trajectories $\braket{\bm{\theta}^{(k)}|\hat{H}|\bm{\theta}^{(k)}}$ against $t=k\eta$ for 30 realizations of the SGD from a certain initial condition.
The learning rate $\eta$ is 0.05 and the number of measurement shots at each step is 100.
As shown, the SGD dynamics often shows a plateau, which is due to a saddle point.

In a plateau of \cref{fgr:SGDsaddle}, the parameters stay near a saddle point with the energy around $-6.470$.
At that saddle, the Hessian has eigenvalues
$(8.76$, $7.77$, $6.68$, $5.99$, $4.44$, $4.15$, $3.46$, $1.97$, $1.10$, $0.358$, $0.00850$, $0.00397$, $0.00$, $-0.0112$, $-0.0337$, $-0.118)$.

We measure the escape time from this saddle point.
The initial state is chosen as the state at $t=5.4$ generated by the GD in \cref{fgr:SGDsaddle}, i.e., near the saddle point mentioned above.
Starting from this initial state $\bm{\theta}^{(0)}$, we update the parameters $\bm{\theta}^{(k)}$ via \cref{eq:SGD}.
The escape time $t_\mathrm{esc}$ is identified as the time $t_\mathrm{esc}=\eta k$ at which the energy $L(\bm{\theta}^{(k)})$ falls below $-7.0$.

\begin{figure}[tb]
\centering
\begin{tabular}{c}
(a) Escape time from a saddle point \\
\includegraphics[scale=0.55]{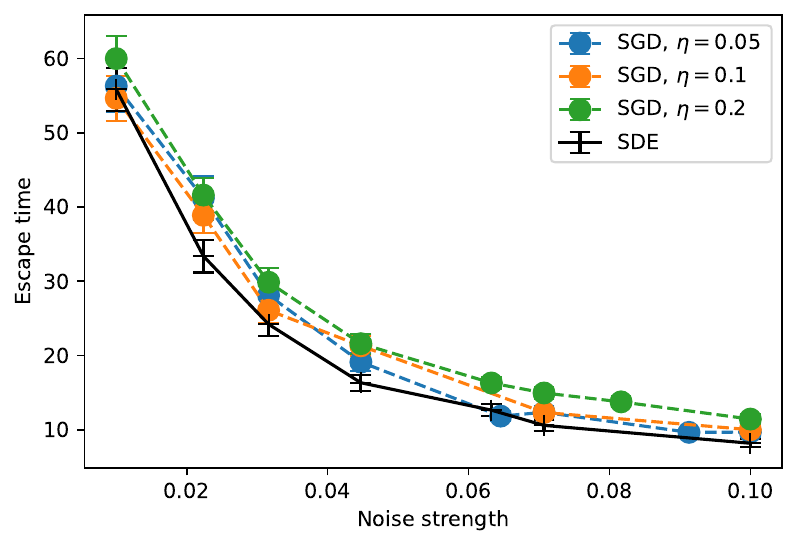}\\
(b)log-log plot\\
\includegraphics[scale=0.55]{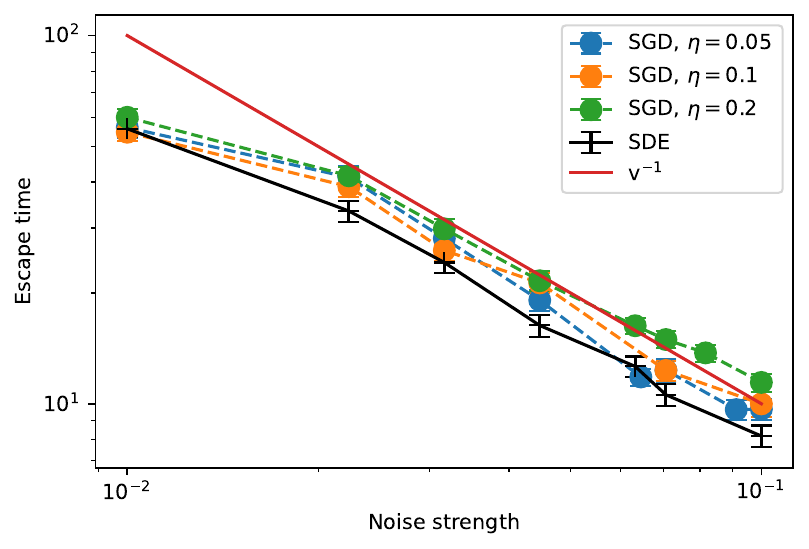}
\end{tabular}
\caption{Escaping time from saddle points with measurement noise.
Both for SGD and the continuous-time SDE, we prepare 100 instances starting from the same initial condition, and compute the escape time averaged over those instances.
We find that the escape time is roughly proportional to the inverse of the noise strength.
}\label{fgr:saddle_etime}
\end{figure}

\begin{figure}[tb]
\centering
\begin{tabular}{c}
(a) Escape time from another saddle point \\
\includegraphics[scale=0.5]{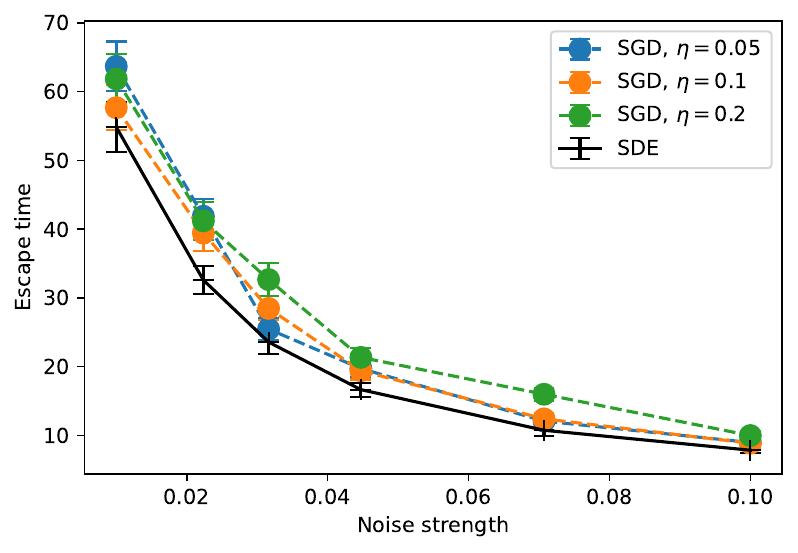}\\
(b) log-log plot\\
\includegraphics[scale=0.5]{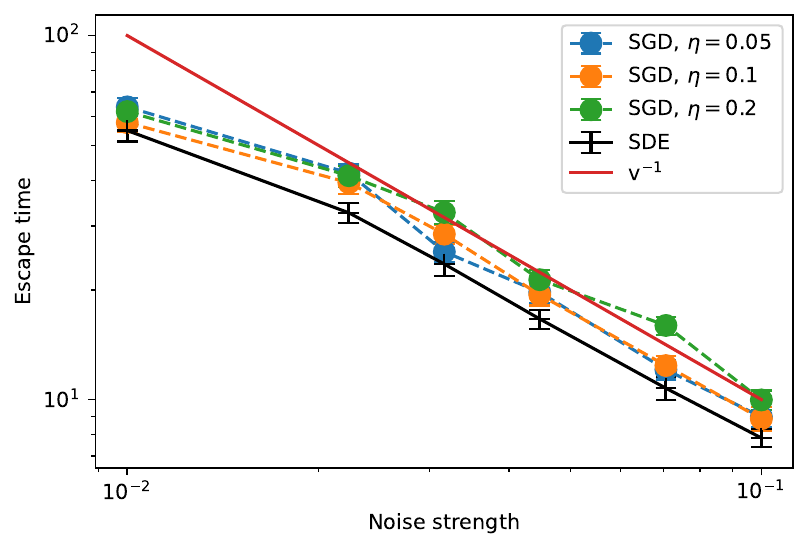}
\end{tabular}
\caption{Escaping time from another saddle point, averaged over 100 instances starting from the same initial condition for SGD with three different values of $\eta$ and SDE.
The escape time is roughly given by the inverse of the noise strength.}
\label{fgr:saddle2_etime}
\end{figure}

\Cref{fgr:saddle_etime} shows escape times from saddle points against $\sqrt{\eta/N_s}$, which turns out to measure the noise strength as discussed in \cref{sec:SDE}, for different values of $\eta$ (black solid lines in \cref{fgr:saddle_etime} show the results for the continuous-time stochastic differential equation introduced later, in \cref{sec:SDE}).
We prepare 100 instances starting from the same initial condition, and the average escape times, as well as its error bars, are plotted.
As intuitively predicted, the escape time decreases as the noise level increases.
Remarkably, the escape times for different values of $\eta$ are concentrated on a single curve when we plot them against $v=\sqrt{\eta/N_s}$.
Furthermore, we find from \cref{fgr:saddle_etime} (b) that the escape time seems to be inversely proportional to the noise strength,
\begin{equation}\label{eq:timesaddle}
t_\mathrm{esc}\propto v^{-1}.
\end{equation}
It should be noted that even the deterministic GD (no noise at all) eventually escapes from the saddle in a finite time, and hence the escape time in the SGD should not diverge in the limit of $v\to +0$.
Therefore, \cref{eq:timesaddle} is valid for not too small values of $v$, and $t_\mathrm{esc}$ should be saturated at a finite value for sufficiently small $v$.

We also study the escape from another saddle point.
Starting from a different initial state, numerical calculations analogous to those in \cref{fgr:SGDsaddle} leads to a different saddle point.
We found a saddle with the energy around $-6.48$ and the Hessian eigenvalues $(9.18$, $7.51$, $6.80$, $6.54$, $5.11$, $3.98$, $2.88$, $1.66$, $1.41$, $0.45$, $0.22$, $-0.019$, $-0.0064$, $0.0018$, $-0.0018$, $0)$.
We evaluated the escape time from this saddle point as shown in \Cref{fgr:saddle2_etime}. 
When plotted against $v=\sqrt{\eta/N_s}$, the escape times for different values of $\eta$ are again concentrated on a single curve.
It suggests that the escape times from a saddle point is generally expressed as a function of $v=\sqrt{\eta/N_s}$. 
Furthermore, we again see that the escape time is inversely proportional to the noise strength, $t_{esc} \propto v^{-1}$.

Here, let us comment on the $\eta$ and $N_s$ dependencies of the overall measurement cost. 
The measurement cost is defined as the total number of quantum measurements required to escape from the saddle.
It is proportional to $N_st_\mathrm{esc}/\eta=t_\mathrm{esc}/v^2$ (it follows from the fact that $t_\mathrm{esc}/\eta$ is the total number of steps to escape the saddle).
Since $t_\mathrm{esc}$ is a function only of $v$, the measurement cost is also written as a function of $v$.
Therefore, for a fixed value of the noise strength $v$, the total number of measurements required to escape is the same regardless of whether the learning rate is increased or decreased.
By considering \cref{eq:timesaddle}, we find
\begin{align}
\text{measurement cost}\propto v^{-3}
\end{align}
for not too small values of $v$.
In the limit of $v\to +0$, $t_\mathrm{esc}$ should be finite and hence
\begin{align}
\text{measurement cost}\sim v^{-2} \quad (v\to+0).
\end{align}



\subsection{Escaping from an excited state}
\label{sec:excited}

We find a parameter $\bm{\theta}_\mathrm{exc}$ in which $\ket{\psi(\bm{\theta}_\mathrm{exc})}$ is close to an exact excited state with the energy $-4$.
At $\bm{\theta}_\mathrm{exc}$, the gradient of the loss function almost vanishes, and the eigenvalues of the Hessian are ($7.84$, $6.19$, $5.58$, $4.50$, $3.95$, $3.49$, $1.36$, $0.982$,
 $0.574$, $0.395$, $0.272$, $5.78\times 10^{-5}$, $0.00$, $-3.20\times 10^{-7}$, $-1.54\times 10^{-5}$, $-8.28\times 10^{-5}$).
So $\bm{\theta}=\bm{\theta}_\mathrm{exc}$ corresponds to a marginally stable point, where the minimum eigenvalue of the Hessian is very close to zero (i.e. the loss landscape has some flat directions).

The escape time $t_\mathrm{esc}$ is now defined as time when the energy value falls below $-5.0$.

\begin{figure}[tb]
  \centering
  \begin{tabular}{c}
  (a) Escape time from an excited state \\
  \includegraphics[scale=0.55]{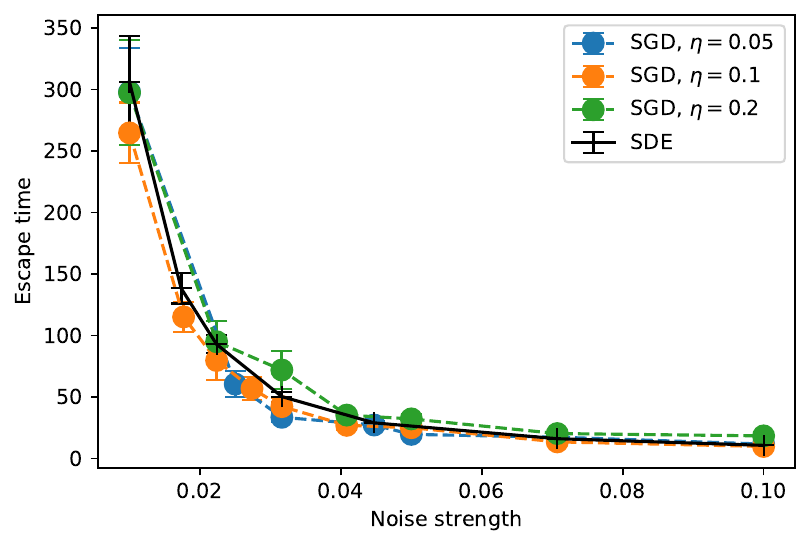}\\
  (b) log-log plot\\
  \includegraphics[scale=0.55]{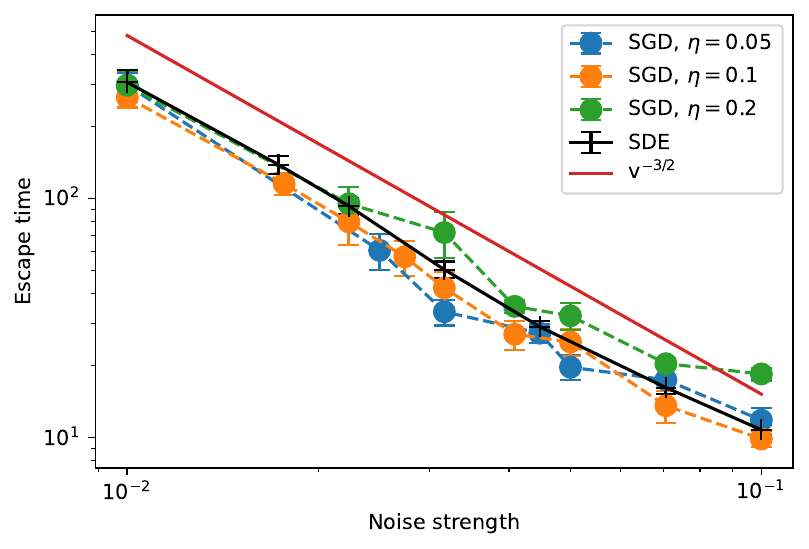}
  \end{tabular}
  \caption{Escape times from an excited state, averaged over 100 instances starting from the same initial condition, for SGD with three different values of $\eta$ and the continuous-time SDE. We find the escape time behaves as $v^{-3/2}$ as indicated by the red line.}\label{fgr:LM_etime}
\end{figure}

\Cref{fgr:LM_etime} shows the escape time from the excited state. We prepare 100 instances starting from the same initial conditions.
We also investigate the escape time from excited, as depicted in \Cref{fgr:LM_etime}. Similarly to \Cref{fgr:saddle_etime} and \Cref{fgr:saddle2_etime}, when the escape times are plotted against $v=\sqrt{\eta/N_s}$, we observe that they converge onto a single curve. This analysis further supports the notion that the escape times from the saddle points and excited state are characterized by a function of the noise strength $\eta/N_s$.
Furthermore, we find that the escape time behaves as
\begin{equation}\label{eq:timeLM}
t_\mathrm{esc}=v^{-3/2},
\end{equation}
which differs from the scaling in the escape from a saddle (see \cref{eq:timesaddle}).

\section{Validity of Continuous time approximation for SGD in VQE}
\label{sec:SDE}

When the parameter update for each iteration is small, which is typically the case when the learning rate $\eta$ is small enough, we can consider the continuous-time approximation\cite{li2017stochastic,smith2017bayesian}.
When analyzing discrete SGD theoretically, various studies have been conducted using stochastic differential equations with a continuous-time approximation~\cite{mandt2015continuous,Li2015Dynamics,marceau2017natural,hu2017diffusion,chaudhari2018stochastic,zhu2018anisotropic,An_2019}. 
However, Yaida~\cite{yaida2018fluctuation} has argued that the properties of steady-state fluctuations in discrete SGD and continuous SDE are different.
Therefore it is important to verify the validity of the continuous-time approximation of SGD in VQE.

\subsection{Stochastic differential equation for SGD in VQE}\label{sec:SDE_SGD}

For small enough constant learning rate $\eta$, \cref{eq:SGD} is regarded as a discrete update (i.e. the Euler-Murayama discretization) of the following SDE by considering $\eta$ as an infinitesimal time step: $\eta=dt$~\citep{mandt2015continuous,li2017stochastic,smith2017bayesian}:
\begin{equation}\label{eq:SDE}
d\bm{\theta}_t=-\bm{\nabla}L(\bm{\theta}_t)dt +\sqrt{\eta C(\bm{\theta}_t)}\cdot dW_t,
\end{equation}
where $W_t$ is a Wiener process, and $dW_t=\mathcal{N}(0,I_pdt)$ with $I_n$ being the $n$-by-$n$ identity matrix. The product $\sqrt{\eta C(\bm{\theta}_t)}\cdot dW_t$ is interpreted as It\^{o}.
The continuous time variable $t$ corresponds to $\eta k$ in the discrete SGD.


A tricky point is that we have $\eta$ dependence of the noise term in \cref{eq:SDE}.
This remaining $\eta$ should be treated as a finite value, although we take the limit of $\eta\to +0$ in the derivation of \cref{eq:SDE}.
Such a treatment is rather heuristic and inconsistent from the mathematical viewpoint.
It means that the validity of the SDE is not necessarily ensured even for small $\eta$, and hence we should care about the validity of the SDE.
This subtlety is further discussed in \cref{sec:properties}, but we find that the SDE correctly reproduces the escape time from saddles and an excited state calculated by VQE in \cref{sec:escape}.
See \cref{fgr:saddle_etime,fgr:saddle2_etime,fgr:LM_etime}.

An immediate implication of \cref{eq:SDE} is that the effect of noise should appear as a function of $v\coloneqq\sqrt{\eta/N_s}$.
Reducing $N_s$ has the same effect as increasing $\eta$.
Therefore, as long as \cref{eq:SDE} is valid, the escape time must be written as a function of $v$, which is a nontrivial scaling relation predicted by the SDE.

\subsection{Properties of steady-state fluctuations in VQA}
\label{sec:properties}
Yaida~\cite{yaida2018fluctuation} derived ``fluctuation-dissipation relations'' for the SGD algorithm.
These relations can be used to adaptively set training schedule and be used to the relations to efficiently extract information pertaining to a loss-function landscape such as the magnitudes of its Hessian and anharmonicity. 
Using the stationarity assumption that the $k$th and $(k+1)$th steps of the optimization process have the same probability distribution, the following relation is derived~\citep{yaida2018fluctuation}:
\begin{equation}\label{eq:FDR1}
\langle({\bm{\theta}}\cdot \nabla L({\bm{\theta}}))\rangle =\frac{\eta}{2}\braket{\Tr\Tilde{C}(\bm{\theta})},
\end{equation}
where stationary-state average is defined as
\begin{equation}
\langle O({\bm{\theta}})\rangle = \int d\bm{\theta}p_{ss}({\bm{\theta}})O({\bm{\theta}})
\end{equation}
for an arbitrary observable $O({\bm{\theta}})$ and for an arbitrary stationary distribution $p_{ss}(\bm{\theta})$, which satisfies $p_{ss}(\bm{\theta}^{(k)})=p_{ss}(\bm{\theta}^{(k+1)})$.

A remarkable point of \cref{eq:FDR1} is that $\Tilde{C}$ in the right hand side is \emph{not} the noise covaraince matrix $C(\bm{\theta})$ but the second-order moment matrix of the gradient, which is defined as
\begin{equation}\label{eq:moment}
\Tilde{C}_{i,j}(\bm{\theta}) \coloneqq \mathbb{E}[g^{(t)}_i({\bm{\theta}})g^{(t)}_j({\bm{\theta}})].
\end{equation}
If we derive an analogous relation for the continuous-time SDE, i.e., \cref{eq:SDE}, we would obtain the equation in which $\tilde{C}(\bm{\theta})$ is replaced by $C(\bm{\theta})$ in \cref{eq:FDR1}.
This difference stems from the subtlety of the derivation of the SDE explained in \cref{sec:SDE_SGD}.
The above fact suggests that the continuous-time SDE is not valid when $C(\bm{\theta})$ significantly differs from $\tilde{C}(\bm{\theta})$.

It should be noted that \cref{eq:FDR1} is valid for a quasi-stationary distribution that is localized at a saddle point or a local minimum.

\begin{figure}[tb]
  \centering
  \includegraphics[scale=0.55]{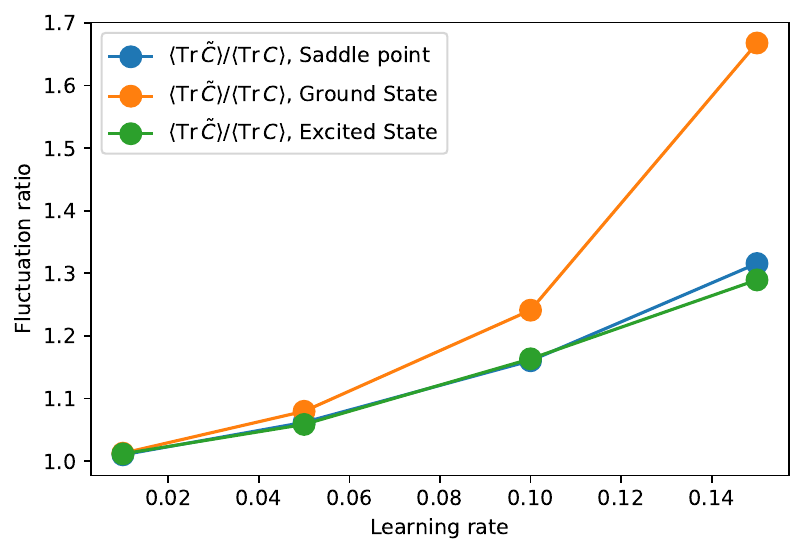}\\
  \caption{Comparison of $\Tilde{C}$ and $C$ at (quasi-)stationary state around the ground state, the saddle point same as in \cref{fgr:saddle_etime}, and the excited state same as in \cref{fgr:LM_etime}. We plot the values of $(\Tr\tilde{C})/(\Tr C)$ for varying $\eta$ with the noise level $\eta/N_s$ fixed to be $0.0005$.} \label{fgr:tildec}
\end{figure}

\Cref{fgr:tildec} shows the ratio of $\braket{C(\bm{\theta})}$ to $\braket{\Tilde{C}(\bm{\theta})}$ as a function of the learning rate $\eta$ at the ground state (a global minimum of the loss function), a saddle point of \cref{fgr:saddle_etime}, and an excited state of \cref{fgr:LM_etime}.
We compute $\braket{C(\bm{\theta})}$ and $\braket{\tilde{C}(\bm{\theta})}$ by considering a time average instead of the average over a (quasi-)stationary distribution $p_{ss}$:
\begin{align}
    \braket{O(\bm{\theta})}\approx\frac{1}{n}\sum_{k=1}^n O(\bm{\theta}^{(k)}),
\end{align}
where $\bm{\theta}^{(k)}$ is generated by \cref{eq:SGD} starting near a critical point (the ground state, a saddle point, or an excited state) and the time step $n$ is chosen as large as possible but smaller than the escape time step.
\Cref{fgr:tildec} shows that, by changing $\eta$ with $\eta/N_s$ held fixed, $C$ and $\tilde{C}$ are close to each other for sufficiently small learning rate, but they show strong deviations by increasing $\eta$.
In this way, fluctuations at a steady state in the SDE differ from those in the SGD unless the learning rate is small enough.
However, \cref{fgr:saddle_etime,fgr:saddle2_etime,fgr:LM_etime} show that this difference does not significantly affect the escape time.

\section{Experimental evaluation on real quantum devices}
\label{sec:dev}
\begin{figure}[tb]
  \centering
  \includegraphics[scale=0.6]{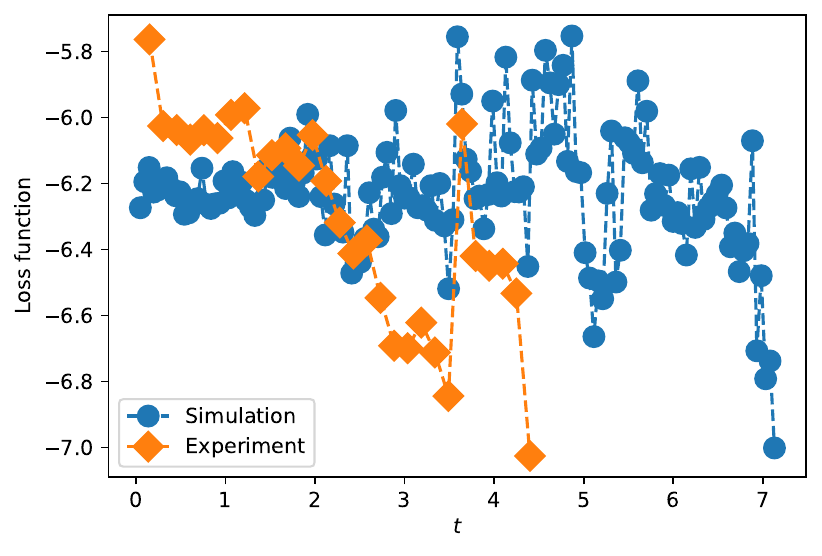}\\
  \caption{Time evolutions of the averaged loss function starting from the saddle point same as in \cref{fgr:saddle_etime}. Blue circles show the result for QASM Simulator averaged over 20 instances, and orange diamonds show the experimental result for ibm\_perth and ibm\_lagos quantum device of the IBM Quantum Systems averaged over 5 instances.
  In both cases, we set $\eta=0.2$ and $N_s=20$. }\label{fgr:device}
\end{figure}

We evaluated the escape time from the saddle point on near-term quantum devices and compared it with the simulation results.
We use 4-qubit Hamiltonian (\cref{eq:hamiltonian}) on the 7-qubit ibm\_perth and ibm\_lagos quantum devices of the IBM Quantum Systems.
\Cref{fgr:device} shows the evolution of the loss function averaged over 5 instances starting from the saddle point considered in \cref{fgr:saddle_etime} on quantum devices of the IBM Quantum Systems (blue circles).
We compare it with the evolution of the loss function averaged over 20 instances for Qasm Simulator (orange diamonds). 
In both cases, we set $\eta=0.2$ and $N_s=20$. 

It is difficult to give a decisive conclusion based on only 5 instances, but we find a tendency of faster escape in the real quantum device compared with the classical simulator.
It would be due to the fact that the real quantum device suffers from additional noise, which may help getting out of the saddle point, but this is a future issue for the effect of more general noise sources in the real quantum devices.

\section{Conclution}
\label{sec:conclusion}

We analyzed the effect of measurement noise, i.e., the finite number of measurements, on the implementation of the gradient-based methods on an actual quantum computer.
One of the main finding is that the covariance matrix of measurement shot noise has no off-diagonal matrix element: $C_{ij}(\bm{\theta})=0$ for any $i\neq j$. It is in stark contrast to the SGD noise in classical machine learning, which arises from mini-batch sampling. The measurement noise of SGD in VQE is given by a full two-point noise matrix $\Tilde{C}$ for fluctuations in ground states and stationary states such as saddle points and excited states. On the other hand, the steady-state fluctuations of SDE are given by two-point noise matrix $C$. Therefore, the properties and structure of the fluctuations are different for VQE and SGD when the learning rate is large. However, the continuous-time approximation of SDE is well established and this difference does not affect the escape from the plateau. 
This implies that the behavior of finite-shot SGD can be interpreted using SDE analysis.

We also found that the computational cost required to escape from a plateau faster does not depend on the learning rate or the number of measurements at each step, but on the strength of the noise, which is the ratio of these factors.
Furthermore, we show that there is a power-law relationship in some cases between the escape time from the saddle point or excited state and the intensity of the noise.
We find that the power-law behavior is also observed in another model (see in \cref{app:XYZ}), while the exponent of power is model-dependent.

As for experimental evaluation on the real quantum devices, the escape time from the saddle points tend to be faster on the real quantum devices, although the number of trials is small, indicating that the real device noise may help the escape from the saddle points.
It will be interesting to see how the entanglement structure of the ansatz and the dimensionality of the parameters affect the escape from the plateau, as well as the effect of the devices noise, but this is a topic for future work.

\begin{acknowledgments}
E.K was supported by JSPS Grant Number 20K14388 and JST PRESTO Grant number JPMJPR2011.
T. M was supported by JST PRESTO Grant number JPMJPR2259.
E.K and T. M were supparted by JSPS Grant Number 21H05185.
In addition, E.K., M.S. and N.Y. were supported by the MEXT Quantum Leap Flagship Program Grant Number JPMXS0118067285 and JPMXS0120319794. We would like to thank  Dr.~Yutaka~Shikano and Dr.~Tatsuhiko~Shirai for technical discussion.
\end{acknowledgments}

\bibliography{references}

\clearpage

\onecolumngrid
\appendix
\section{Appendix}
\subsection{Escaping from saddle points in XYZ model}\label{app:XYZ}
We also consider the problem of finding the ground state of the one-dimensional (1D) XYZ Heisenberg model.
The Hamiltonian is given by
\begin{equation}\label{eq:XYZhamiltonian}
H=\sum_{i=1}^4~J_xX_iX_{i+1}+J_yY_iY_{i+1}+J_zZ_iZ_{i+1}.
\end{equation}
Here, $J_x=1.421$, $J_y=1.288$, $J_z=1.0$.
The ansatz has the same settings as in \cref{sec:saddle}, which is a typical hardware-efficient ansatz illustrated in Fig. \ref{fgr:ansatz}.
We measure the escape time from this saddle point.
The parameters stay near a saddle point with the energy around $-9.43$.
The eigenvalues of the Hessian at the saddle are (13.30, 10.31, 9.64,   8.38,  6.98, 4.87, 4.31,  3.23,  2.67,  2.54,  0.93,   0.37, 0.25, -0.16, -0.06,  0.052).
The escape time is determined by the point in time when the energy $L(\bm{\theta}^{(k)})$ falls below $-9.7$.
Based on the continuous-time SDE discussed in \cref{sec:SDE}, the noise strength of SGD is defined as $v=\sqrt{\eta/N_s}$.
Similarly in \cref{sec:escape} and \cref{sec:SDE}, the escape time from the saddle point for different values of $\eta$ are concentrated on the single curve when we plot them against $v=\sqrt{\eta/N_s}$.

The escape time behaves as
\begin{equation}
t_\mathrm{esc}\propto v^{-1/2}.
\end{equation}
Although the exponent of power differs from that in the isotropic Heisenberg model discussed in the main text, there is a power-law relation between escape time and noise strength.

\begin{figure}[h]
  \centering
  \begin{tabular}{cc}
  (a) Escape time from a saddle point &
  (b) Log-log plot \\
  \includegraphics[keepaspectratio, scale=0.5]
  {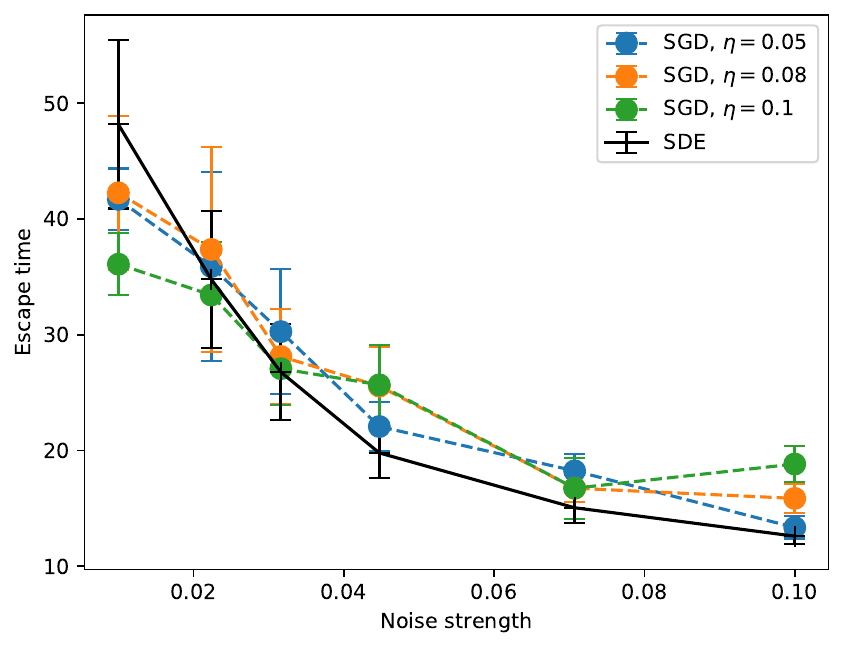}&
  \includegraphics[keepaspectratio, scale=0.5]
  {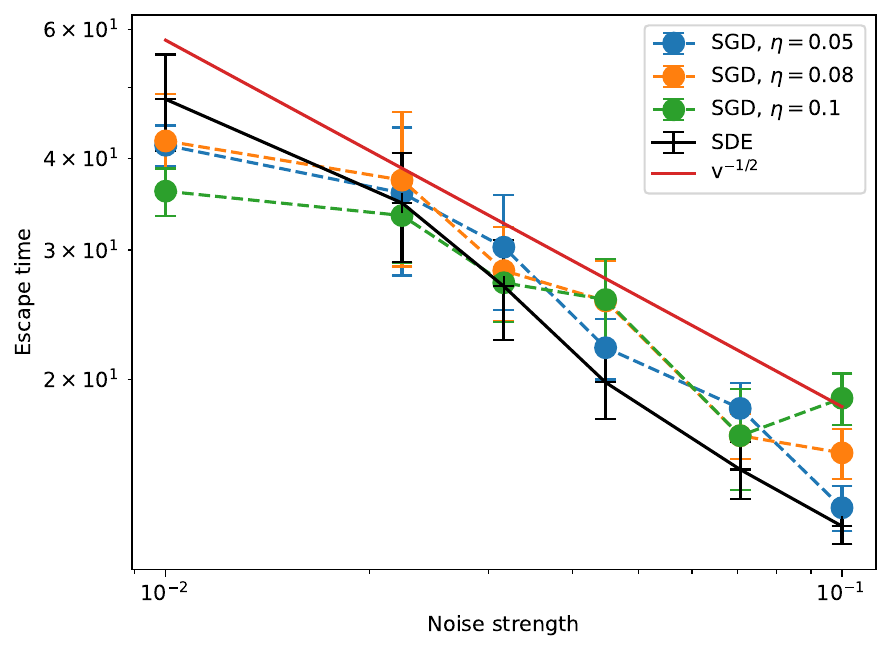}
  \end{tabular}
 \caption{Escape time from a saddle point. We prepare 100 instances starting from the same initial condition, and compute the escape time averaged over them. The continuous-time SDE provides a good approximation even when the learning rate is relatively large.}\label{fgr:XYZ}
\end{figure}

\end{document}